\documentclass[reprint,twocolumn,aps,prl,showpacs, showkeys,amsmath,amssymb, groupaddress]{revtex4}

\usepackage{graphicx}
\usepackage{dcolumn}
\usepackage{bm}

\begin{document}


\title{Coalescence of Pickering emulsion droplets induced by electric-field}
\author{Guo Chen$^1$, Peng Tan$^{1}\footnote{E-mail: tanpeng0928@gmail.com}$, Shuyu Chen$^2$, Jiping Huang$^3$, Weijia Wen$^2$ and Lei Xu$^{1}\footnote{E-mail: xulei@phy.cuhk.edu.hk}$}
\affiliation{$^1$Department of Physics, The Chinese University of Hong Kong, Hong Kong, China\\
$^2$Department of Physics, The Hong Kong University of Science and Technology, Hong Kong, China\\
$^3$ Department of Physics, Fudan University, Shanghai, China}

\begin{abstract}
Combining high-speed photography with electric current measurement,
we investigate the electrocoalescence of Pickering emulsion
droplets. Under high enough electric field, the originally-stable
droplets coalesce via two distinct approaches: normal coalescence
and abnormal coalescence. In the normal coalescence, a liquid bridge
grows continuously and merges two droplets together, similar to the
classical picture. In the abnormal coalescence, however, the bridge
fails to grow indefinitely; instead it breaks up spontaneously due
to the geometric constraint from particle shells. Such
connecting-then-breaking cycles repeat multiple times, until a
stable connection is established. In depth analysis indicates that
the defect size in particle shells determines the exact merging
behaviors: when the defect size is larger than a critical size
around the particle diameter, normal coalescence will show up; while
abnormal coalescence will appear for coatings with smaller defects.

\end{abstract}


\maketitle

When two droplets come into contact, they naturally coalesce to
minimize the surface energy, a phenomenon extensively studied since
19th century \cite{thomson, duchemin, aartsjfm, caseprl, casepre}. A
quite recent study reveals that coalescence starts from the regime
controlled by inertial, viscous, and surface-tension forces
\cite{pnas}, which is followed by either a viscous regime
\cite{aartsprl} or an inertial regime \cite{zaleski}. However, the
coalescence of special droplets -- Pickering emulsion droplets --
remains poorly understood. Stabilized by colloidal particles instead
of surfactant molecules, Pickering emulsions are composed by
particle-coated droplets \cite{Ramsden, pickering}, as shown in
Fig.1a. Due to the highly controllable permeability, mechanical
strength and biocompatibility \cite{velev, dinsmore, lee}, Pickering
emulsions have been actively studied in the last decade, and may
find broad applications in important areas such as oil recovery
\cite{eow} and drug-delivery \cite{simovic}. The well-controlled
coalescence in Pickering emulsions can also facilitate material
mixing and benefit the field of chemical and biochemical assays
\cite{Ahn}. More interestingly, the existence of an extra structure
-- particle shell -- may bring fundamentally different merging
physics and enrich the classical coalescence research. Consequently,
there is great scientific and practical significance to clarify the
coalescence of Pickering emulsion droplets.

If the surface is poorly coated, droplets can coalesce spontaneously
and form supracolloidal structures \cite{Subramaniam, andrea}.
Complex dynamics and structure of particles are observed during
coalescence, due to the combined effects of charge, surface tension
and liquid flow \cite{fuller}. Numerical simulation further reveals
that the repulsion between particles, the particles' ability to
attach to both droplet surfaces, and the stability of the liquid
film between droplets are crucial for coalescence behaviors
\cite{alberto}. However, if the surface is coated by closely packed
particles, coalescence rarely occurs. Inspired by the strong
influence of electric field, which can deform liquid surface into
conical tips \cite{taylor, brazier, stone, basaran}, create liquid
jets out of these tips \cite{basaran}, induce electrocoalescence
between two isolated droplets \cite{brazier} or among a large
population of droplets \cite{basaran2}, and even prevent droplets
from coalescing \cite{bird, ristenpart}; we thus apply high voltage
between two Pickering emulsion droplets and investigate their
coalescence under electric field. The originally-stable droplets
coalesce systematically, via two distinct approaches: normal
coalescence and abnormal coalescence. In the first approach, a
liquid bridge grows continuously and merges two droplets together,
similar to the classical picture. In the second approach, however,
completely different behaviors emerge: a liquid bridge forms through
defects but fails to grow indefinitely; instead it breaks up
spontaneously due to the geometric constraint from particle shells.
Such connecting-then-breaking cycles repeat multiple times, until a
stable connection is established. Further analysis indicates that
the defect size in particle shells determines the exact merging
behaviors: when defects are larger than a critical size, the normal
approach will show up; while the abnormal coalescence will occur for
smaller defects. This study generalizes the understanding of
coalescence to a more complex system.

\begin{figure}
\includegraphics[width=3.2in]{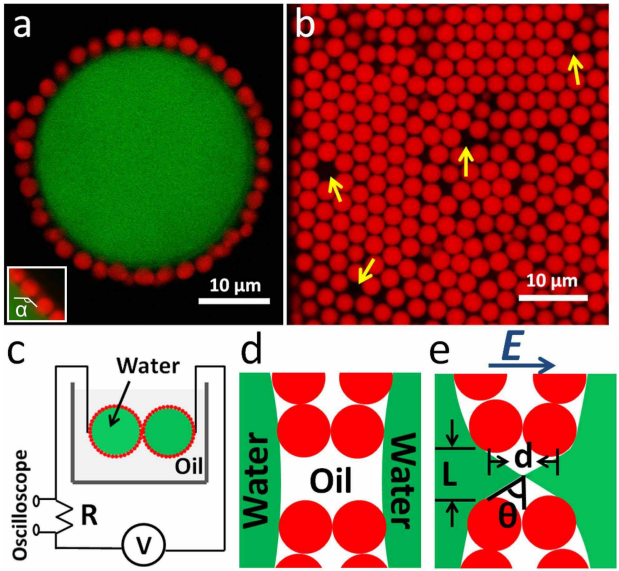}
\caption{(color online). (a) The 2D confocal slice of a typical
Pickering emulsion droplet. The saltwater droplet is dyed in green
and the PMMA particles are dyed in red. The particles have the
diameter $d=3\mu m$ for this particular sample. The magnified image
in the inset shows the contact angle of particles at the interface:
$\alpha=133\pm5^\circ$. (b) The arrangement of particles in the
shell. The yellow arrows point to some typical defects. (c) Diagram
of the experimental setup. A DC voltage is applied through two
electrodes that are in direct contact with the droplets. The current
between two droplets is recorded by an oscilloscope. (d) A cartoon
illustrating the separation of water droplets without electric
field. (e) With high enough electric filed, a liquid bridge is
formed through defects. $d$ is particle diameter, $L$ is the defect
size, and $\theta$ is the cone angle.}
\end{figure}

We make Pickering emulsions by suspending saltwater droplets (12.5
wt\% NaCl) in the organic solvent decahydronaphthalene (DHN). The
droplets are stabilized by Poly(methyl methacrylate) (or PMMA)
particles of diameter $d$. To ensure the general robustness of the
results, we vary $d$ from $0.3\mu m$ to $3\mu m$ in the experiment.
The emulsion droplets have the typical size of a few millimeters.
One such droplet is illustrated by confocal microscopy in Fig.1a,
with the internal saltwater fluorescently labeled in green and the
PMMA particles dyed in red. Apparently, particles coat the droplet
surface, forming a protective shell \cite{velev, dinsmore}. The
magnified image in the lower inset shows the contact angle of
particles at the interface: $\alpha=133\pm5^\circ$. Fig.1b focuses
on the particle shell: most particles are closely packed in the
shell, with some defects randomly distributed \cite{irvine}, as
indicated by the arrows. These defects play an important role in
coalescence, as demonstrated later by our study.

We directly visualize the merging of millimeter-size droplets with a
fast camera (Photron SA4), at the frame rate of $10,000s^{-1}$. A DC
electric voltage is applied through two electrodes that are in
direct contact with the droplets (see Fig.1c). We gradually increase
the voltage until coalescence takes place. To probe microscopic
events the fast camera can not resolve, we further measure the
electric current between the two droplets with an oscilloscope
\cite{caseprl, casepre}. The current measurement is sensitive to any
tiny connection at the beginning of coalescence. The time resolution
of this approach is determined by the relaxation time of charging
process in our salt solution and has the typical value of
$\tau\sim10^{-10}s$ \cite{footnote}. These two approaches
(high-speed photography and electric current) are synchronized to
reveal a complete picture at both the macroscopic and the
microscopic levels. The sketch of the system is shown in Fig.1c.

When electric field is absent or small, two contacting droplets
never coalesce in our experiment, due to the protection of particle
shells. However, sufficiently strong electric field can induce
conical tip structures at the droplet surface \cite{taylor, basaran,
stone}, which may penetrate through the large pores of defects and
form a connection, as demonstrated by the cartoon in Fig.1e. To
simplify the analysis and illustrate the underlying physics, we
assume that the defects in the two shells have identical size and
align perfectly. We define the particle diameter as $d$, the defect
size as $L$, and the the cone angle as $\theta$ \cite{bird}. The
electric field applies a local electrical stress, $\Sigma_E\sim
\epsilon \epsilon_0 E_{loc}^2$, onto the interface. Once $\Sigma_E$
exceeds the restoring capillary pressure, $P_{cap}\sim\gamma/L$,
significant deformation will occur and a liquid bridge may form.
Here $\epsilon \epsilon_0$ is the permittivity of the oil, $E_{loc}$
is the local electric field around the defect, and $\gamma$ is the
surface tension of the water-oil interface. Apparently, the local
electric field must exceed a finite value,
$E_{loc}\sim\sqrt{\gamma/(\epsilon \epsilon_0 L)}$, to establish a
connection.

\begin{figure}
\includegraphics[width=3.2in]{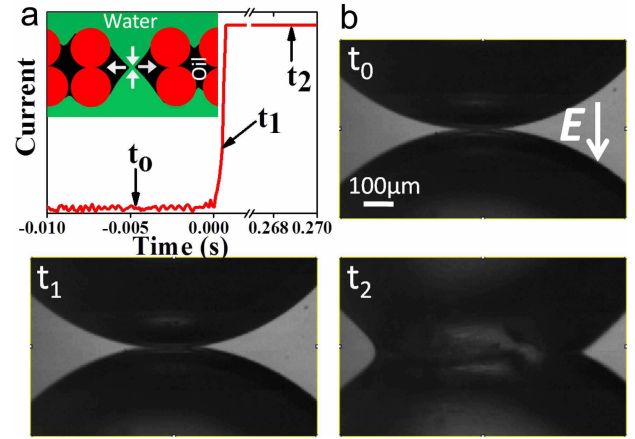}
\caption{(color online). (a) The electric current between two
droplets during a normal coalescence. Before connection, the signal
fluctuates about zero (see $t=t_0$). Around $t=t_1$, however, it
rises rapidly and saturates the apparatus, revealing the
establishment of a stable connection. Inset cartoon illustrates the
widening of the connection by the water flow. (b) The corresponding
images from the synchronized fast photography. At $t_0=-4.7ms$, the
droplets are intact. At $t_1=0.5ms$, although a connection is
suggested by the electric signal, its small size prevents the direct
observation. Significant variation in profile only occurs long after
the connection, as illustrated at $t_2=268.8ms$. The voltage between
two droplets is $50V$ for this particular event.}
\end{figure}

Once connected, the liquid bridge may continuously widen and merge
the two droplets, as illustrated by the cartoon of Fig.2a (see the
supplemental movie S-1 for real situation). The main panel of Fig.2a
plots the electric current signal from oscilloscope. At $t=t_0$, no
connection is established and the signal fluctuates about zero.
Around $t=t_1$, however, the signal rises sharply and saturates the
apparatus, indicating the formation of a stable connection. We
define the starting point of this sharp rise as time zero throughout
our measurements. The three typical moments, $t_0=-4.7ms$,
$t_1=0.5ms$, and $t_2=268.8ms$, are illustrated by the high-speed
images in Fig.2b. Clearly, the connection at $t_1$ barely changes
the macroscopic picture, showing the limitation of photography and
the essence of electrical measurements. Significant change in
profile is only observed long after the connection, as demonstrated
by the image at $t_2=268.8ms$.

The entire process lasts several hundred milliseconds, much longer
than the water droplet coalescence without particle shells
($\sim1ms$) \cite{aartsprl}. We also notice that coalescence may
stop in the middle and never reach the state shown in the $t_2$
image. These behaviors are caused by the particle shells, within
which jammed particles can slow down or even terminate coalescence
\cite{andrea}.

More interestingly, a completely different merging process --
abnormal coalescence -- is observed. The behaviors of the electric
signal is shown in Fig.3a: instead of increasing monotonically to
saturate the oscilloscope, the signal exhibits multiple peaks, as
demonstrated by the ones at $t_1$ and $t_3$. Fig.3b plots the
zoomed-in details of one particular peak, and Fig.3c shows the
corresponding images from synchronized photography.

The high-speed images reveal the detailed dynamics during abnormal
coalescence: due to electrical attraction, the two droplets approach
each other, flattening the contact area but failing to merge, as
shown by the $t_0$ image in Fig.3c. In the subsequent development,
however, the two droplets repeatedly approach then move away from
each other, making multiple oscillatory-like cycles (see the
supplemental movie S-2). Surprisingly, these cycles coincide exactly
with the major peaks in the electric signal, as illustrated in
Fig.3b: the rapid-rising of signal around $t_2$ corresponds to the
approaching of droplets, the local maximum at $t_3$ matches the
moment of closest distance between two droplets, and the
sharp-declining part around $t_4$ agrees with the moving away of two
droplets. Similar cycles repeat multiple times, until a stable
connection is reached, as illustrated in the circle of the $t_5$
image. For this particular experiment, coalescence stops at this
point, due to the particle jamming inside the shells.

These observations suggest the following picture for each cycle: the
electric field deforms the droplet surfaces and establishes a
connection through large pores, as illustrated in Fig.1e. This
connection then quickly broadens, increasing the electric signal and
pulling two droplets together. However, after reaching a maximum
thickness, the bridge shrinks and breaks up, reducing the current to
zero value; correspondingly the two droplets move away from each
other, possibly due to the elastic repulsion from particle shells.

\begin{figure}
\includegraphics[width=3.2in]{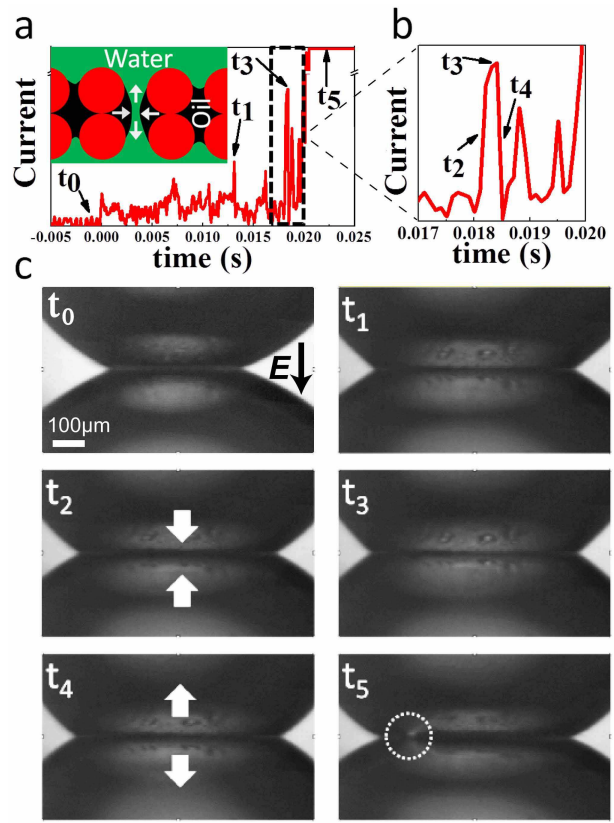}
\caption{(color online). (a) The electric current signal for an
abnormal coalescence. From no connection at $t_0$ to the stable
connection at $t_5$, the signal exhibits multiple cycles of peaks,
such as the ones at $t_1$ and $t_3$. The inset cartoon illustrates
the narrowing of the neck due to the outward water flow, which
breaks the bridge and completes one peak cycle. (b) The zoomed-in
picture for one particular peak at $t_3$. The signal rises sharply
around $t_2$, reaches maximum at $t_3$, and drops dramatically
around $t_4$. (c) The corresponding images from the synchronized
fast photography. At $t_0$, there is no connection but the contact
area is flattened due to the electric attraction. The image at $t_1$
corresponds to one peak in the electric signal. Images $t_2$ through
$t_4$ clarify the particular peak shown in (b). Around $t_2$, the
droplets approach each other as indicated by the arrows, and
corresponds to the sharp rise of the electric signal. At $t_3$, two
droplets are at the closest distance and the signal reaches the
maximum. Around $t_4$, the droplets move away from each other, and
agrees with the dramatic drop of signal. The image of $t_5$ shows a
stable connection inside the circle. The specific times are:
$t_0=1.3ms, t_1=13.1ms, t_2=18.0ms, t_3=18.4ms, t_4=18.6ms$, and
$t_5=22.2ms$. The voltage between two droplets is $80V$.}
\end{figure}

The direct visualization of liquid bridge would serve as an ideal
confirmation for the above picture. However it typically occurs
within the contact area and eludes such observation. After a number
of trials, nevertheless, we successfully catch the cycles of bridge
formation and break-up with fast camera, as shown in Fig.4a: no
connection exists at $t_0$, while a tiny bridge is formed at $t_1$
(inside the box), which subsequently breaks up at $t_2$. Comparing
with the electric signal in Fig.4b, we find an exact correspondence
between the moment of bridge formation in Fig.4a and the peak
location in Fig.4b at $t=t_1$. Similar cycles appear for many times
(see the supplemental movie S-3 for a clear demonstration), and the
correspondence between bridge formation and signal peak is always
observed. These findings unambiguously verify our picture of the
abnormal coalescence.

\begin{figure}
\includegraphics[width=3.0in]{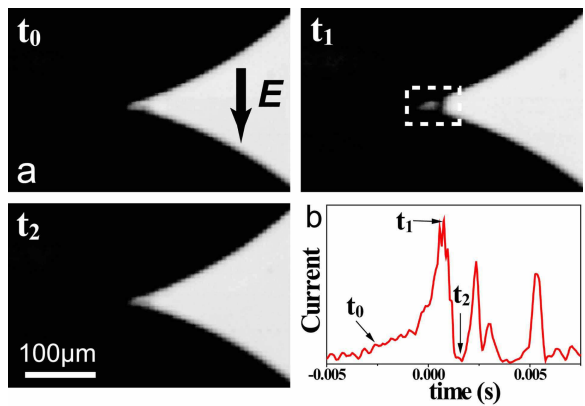}
\caption{(color online). (a) The direct visualization of a liquid
bridge. No connection exists at $t_0=-2.5ms$, while a tiny bridge is
formed at $t_1=1.0ms$ (inside the box), which subsequently breaks up
at $t_2=2.0ms$. (b) The corresponding electric signal. Clearly the
bridge formation in (a) coincides exactly with one particular peak
in the electric signal.}
\end{figure}

One essential question remains unclear: why does the bridge shrink
and break up, instead of growing continuously? We propose a
geometric explanation similar to the non-coalescence mechanism of
charged water droplets \cite{bird, ristenpart}. Because of particle
shells, the droplet surfaces are separated by a fixed distance
around the particle diameter, $d$. Thus the geometry of the bridge
depends mainly on the defect size, $L$ (see the schematics of
Fig.1e). When $L$ is small, only slender cones with large $\theta$
can form, making the capillary pressure from the positive curvature
(the one encircling the neck) dominate the pressure from negative
curvature (the one along the bridge). This positive pressure
consequently pushes liquid back into droplets and breaks the neck
\cite{bird, ristenpart}, leading to abnormal coalescence. By
contrast, however, when the defects are large, cones with small
$\theta$ will form, making the negative curvature dominate the
positive one. As a result, the negative pressure at the neck will
suck more liquid from droplets and widen the connection, causing
normal coalescence. The liquid flow and bridge evolution of the two
situations are illustrated by the cartoons in Fig.2a and Fig.3a
respectively.

Our analysis naturally leads to a critical defect size, $L_c$,
above/below which the normal/abnormal coalescence occurs. This
critical size should correspond to the critical angle,
$\theta_c=30.8^\circ$, in the coalescence to non-coalescence
transition between water droplets \cite{bird}. According to Fig.1e,
we have $\tan\theta\sim d/L$, which gives rise to: $L_c\sim
d/\tan\theta_c\sim1.68d$. Therefore the critical defect size is
around the diameter of a particle, crossing which distinct merging
phenomena will appear. Defects of this size are commonly observed,
as demonstrated in Fig.1a. We further compare the two competing
stresses, the electrical stress ($\epsilon_0\epsilon E_{loc}^2$) and
the restoring capillary pressure ($\gamma/L_c$), at this critical
size. Plugging in the values $\gamma\sim30mN/m$, $\epsilon\sim2$,
and $E_{loc}\sim V/d$ ($V=80v$ and $d=3\mu m$ for the particular
experiment shown in Fig.3), we obtain their ratio,
$\epsilon_0\epsilon E_{loc}^2/(\gamma/L_c)\sim2$. This
order-of-unity ratio confirms again that coalescence only occurs
after the electrical stress overcomes the restoring stress, as
predicted by our model.

\begin{figure}
\includegraphics[width=3.2in]{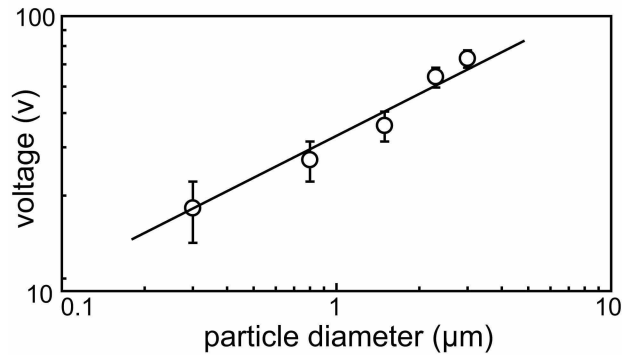}
\caption{The voltage at which oscillation starts to occur, $V$,
versus the particle diameter, $d$. Apparently $V$ increases with
$d$, indicating higher voltage is needed for the oscillation of
droplets coated with larger particles. The dependence is consistent
with the solid line of $V\propto d^{0.5}$ predicted by our model.}
\end{figure}

Moreover, this analysis predicts a general trend for the oscillation
of droplets coated by different sized particles. When the electrical
stress balances the capillary pressure, $\epsilon_0\epsilon
E_{loc}^2\sim(\gamma/L)$, oscillation should start to occur.
Plugging in the typical conditions of $L\sim d$ and $E_{loc}\sim
V/d$, we obtain $V \propto d^{0.5}$. Therefore, for droplets coated
by larger particles, oscillation should start to occur at a larger
voltage $V$. To test this prediction, we perform measurements with
different particle sizes and vary $d$ by one order of magnitude.
Indeed we observe the increase of $V$ with respect to $d$, which is
consistent with the predicted relation $V\propto d^{0.5}$ (see
Fig.5). This agreement provides another experimental evidence for
our model.

In conclusion, we achieve the systematic coalescence of Pickering
emulsion droplets by applying high enough electric field. The
droplets coalesce via two distinct approaches: the normal and
abnormal coalescence. During the normal coalescence, a liquid bridge
grows continuously and merges two droplets together, similar to the
liquid droplet coalescence but with a much slower speed. In the
abnormal coalescence, however, completely different behaviors
emerge: a liquid bridge forms but fails to grow indefinitely;
instead it breaks up spontaneously due to the geometric constraint
of particle shells. This connecting-then-breaking cycle repeats
multiple times, until the defects grow large enough and establish a
stable connection. In depth analysis indicates that when the defect
size is much larger than the particle diameter, normal coalescence
will show up; while the abnormal coalescence appears for smaller
defects. Moreover, droplets coated by larger particles require
higher voltages to oscillate and coalesce. This study generalizes
the understanding of coalescence to a more complex system, and may
find useful applications in the industries related to Pickering
emulsions.

G. C., P. T. and L. X. are supported by Hong Kong RGC (CUHK404211,
CUHK404912 and direct Grant 2060442). J. P. H. is supported by NNSFC
(11075035/11222544), Fok Ying Tung Education Foundation (131008),
and Shanghai Rising-Star Program (12QA1400200).


\begin{thebibliography}{99}
\bibitem{footnote}
  $\tau$ can be calculated as $\tau=\epsilon_{water}\epsilon_0/\sigma$. Here $\epsilon_{water}=86$ is the relative permittivity of water,
  $\epsilon_0=8.85\times10^{-12}F/m$ is the vacuum permittivity, and
  $\sigma=7S/m$ is the conductivity of our salt water. The detailed
  derivation of $\tau$ is the following: from charge conservation we
  have $\nabla\cdot\sigma\overrightarrow{E}+\partial\rho/\partial t=0$; also from Gauss's law we have $\nabla\cdot\epsilon_{water}\epsilon_0\overrightarrow{E}=\rho$.
  Combining the two gives:  $\partial\rho/\partial t + \rho/\tau=0$, with $\tau=\epsilon_{water}\epsilon_0/\sigma$. This equation has the solution: $\rho(t) = \rho_0 \cdot exp(-t/\tau)$,
  which gives the characteristic relaxation time $\tau=\epsilon_{water}\epsilon_0/\sigma\sim10^{-10}s$.
\end{thebibliography}
\end{document}